\documentstyle[preprint,aps,epsfig]{revtex}
\draft
\begin{document}
\preprint{Nucl. Phys. A (2002) in press.}
\title{High Density Behaviour of Nuclear Symmetry Energy
and High Energy Heavy-Ion Collisions}
\bigskip
\author{\bf Bao-An Li\footnote{email: Bali@astate.edu}}
\address{Department of Chemistry and Physics\\
P.O. Box 419, Arkansas State University\\
State University, Arkansas 72467-0419, USA}
\maketitle

\begin{quote}
High energy heavy-ion collisions are proposed as a novel means to 
obtain information about the high density ({\rm HD}) behaviour of 
nuclear symmetry energy. Within an isospin-dependent hadronic transport 
model using phenomenological equations of state ({\rm EOS}) 
for dense neutron-rich matter, it is shown that the isospin 
asymmetry of the HD nuclear matter formed in high energy heavy-ion collisions 
is determined mainly by the HD behaviour of nuclear symmetry energy. 
Experimental signatures in several sensitive 
probes, i.e., $\pi^-$ to $\pi^+$ ratio, transverse collective flow 
and its excitation function as well as neutron-proton differential flow, are 
investigated. A precursor of the possible isospin separation instability in 
dense neutron-rich matter is predicted to appear as the local minima in the 
excitation functions of the transverse flow parameter for both neutrons 
and protons above the pion production threshold. Because of its {\it qualitative} 
nature unlike other {\it quantitative} observables, this precursor can be 
used as a unique signature of the isospin dependence of the nuclear {\rm EOS}. 
Measurements of these observables will provide the first terrestrial data to 
constrain stringently the HD behaviour of nuclear symmetry energy and thus 
also the {\rm EOS} of dense neutron-rich matter. Implications of our findings 
to neutron star studies are also discussed.  
\\
{\bf PACS} numbers: 25.70.-z, 25.75.Ld., 24.10.Lx\\
{\bf Key Words:}High Energy Heavy Ion Collisions, Isospin, Symmetry Energy,
Equation of State, Neutron-Rich Matter, Neutron Stars, Collective Flow
\end{quote}

\newpage
\section{Introduction}
The rapid advance in nuclear reactions using rare isotopes has opened 
up several new frontiers in nuclear 
sciences\cite{tanihata,liudo,pak97,lkb98,sjy,betty,udo,ditoro}. 
In particular, the intermediate energy heavy rare isotopes currently 
available at the National Superconducting Cyclotron Laboratory (NSCL/MSU) 
and the more energetic ones to be available at the future Rare Isotope 
Accelerator (RIA) in the United States provide a unique opportunity to 
explore novel properties of dense neutron-rich matter that was not in reach 
in terrestrial laboratories before. This exploration will reveal crucial
information about the equation of state (${\rm EOS}$) of neutron-rich matter.
To understand the latter and its astrophysical implications, such as,  
the origin of elements, structure of rare isotopes and properties of 
neutron stars are presently among the most important goals of nuclear 
sciences. The ${\rm EOS}$ of neutron-rich matter of isospin asymmetry 
$\delta\equiv (\rho_n-\rho_p)/(\rho_n+\rho_p)$ can be written as
\begin{equation}\label{ieos}
e(\rho,\delta)= e(\rho,0)+E_{sym}(\rho)\delta^2
\end{equation}
within the parabolic approximation (see e.g., \cite{lom}), 
where $e(\rho,0)$ is the energy per nucleon in isospin symmetric 
nuclear matter. To study the density dependence of nuclear symmetry energy 
$E_{sym}(\rho)$ has been a longstanding goal of extensive research with various 
microscopic and/or phenomenological models over the last few decades, e.g., 
\cite{bru67,siemens70,fri81,chin77,serot,har87,mut87,wiringa88,bom91,akm98,hub98,my98,lee98}, 
for a recent review, see, e.g. \cite{lom}. All models are able to give the 
symmetry energy at normal nuclear matter density $E_{sym}(\rho_0)$ of 
about $34\pm 4$ MeV in agreement with that extracted from the atomic mass data. 
However, the predicted results on the density dependence, especially at 
high densities, are extremely diverse and often contradictory. At subnormal 
densities, all models predict that the symmetry energy increase with density 
although the growing rate is rather model dependent. To extract the density 
dependence of symmetry energy at low densities, several approaches have 
recently been proposed. These include measuring the thickness of 
neutron-skins of rare isotopes\cite{oya} and heavy stable nuclei, 
such as $^{208}Pb$\cite{brown,hor,typel,fur}, isospin 
fractionation\cite{lik97,xu,tan} as well as nuclear collective 
flow\cite{ibuu2,sca99,li00} in intermediate energy heavy-ion collisions. 
Above normal nuclear matter densities, even the trend of the density 
dependence of $E_{sym}(\rho)$ is still controversial. Theoretical results 
can be roughly classified into two groups, i.e., a group where 
the $E_{sym}(\rho)$ rises monotonously and one in which it falls with 
the increasing density above about twice the normal nuclear matter density.
The predictions also depend on the nature and form of effective interactions used. 
For instance, within the same Hartree-Fock approach using about 25 
Skyrme and Gony effective interactions that have been widely 
used successfully in studying saturation properties of symmetric nuclear 
matter and nuclear structures near the $\beta$ stability valley, the calculated 
symmetry energies were found to fall approximately equally into the 
two groups\cite{brown,mar}. 

The density dependence of nuclear symmetry energy, 
especially at high densities\cite{kut}, 
has many profound consequences for various studies in 
astrophysics\cite{lat00,pra97,bom01}. In particular, an increasing $E_{sym}(\rho)$ 
leads to a relatively more proton-rich neutron star whereas a decreasing 
one would make the neutron star 
a pure neutron matter at high densities. Consequently, the chemical composition 
and cooling mechanisms of protoneutron stars\cite{lat91,sum94}, critical densities for
Kaon condensations in dense stellar matter\cite{lee96,kurt2}, mass-radius 
correlations\cite{prak88,eng94} as well as the possibility of a mixed 
quark-hadron phase\cite{kurt3} in the cores of neutron stars will all be 
rather different. The high density (HD) behaviour of nuclear symmetry 
energy $E_{sym}(\rho)$ 
is very important for understanding many interesting astrophysical phenomena, 
but it also subjects to the worst uncertainty among all properties of dense 
nuclear matter\cite{kut}. The fundamental cause of the extremely uncertain HD behaviour of 
$E_{sym}(\rho)$ is the complete lack of terrestrial laboratory 
data to constrain directly the model predictions. We explore in this work the
possibility of using high energy heavy-ion collisions to probe the HD behaviour 
of $E_{sym}(\rho)$. A brief report of this work can be found in ref. \cite{li02}. 
The $E_{sym}(\rho)$ is found to affect significantly 
several aspects of high energy heavy-ion collisions. In particular, 
the neutron/proton ratio of HD nuclear matter formed in these collisions 
is determined mainly by the HD behaviour of $E_{sym}(\rho)$.
Several promising signatures of the HD behaviour of $E_{sym}(\rho)$
are also investigated. Besides the {\it quantitative} observables, we
also study one special {\it qualitative} signature of the HD behaviour of the
symmetry energy, i.e., the local minimum in the excitation function of 
nuclear transverse flow. The paper is organized as follows. In Section 2, we 
discuss the equation of state (${\rm EOS}$) for neutron-rich matter, 
in particular, the HD behaviour of nuclear symmetry and its effects on the
neutron/proton ratio in neutron stars. In section 3, we 
study the neutron/proton ratio and its dependence on the symmetry energy 
in HD hadronic matter formed in high energy heavy-ion collisions within 
an isospin-dependent hadronic transport model. In section 4, we explore several
experimental probes of the HD behaviour of nuclear symmetry energy.
Finally, we summarize in section 5.

\section{Equation of state of dense neutron-rich matter}
In this section properties of the equation of state of dense
neutron-rich matter are discussed. In particular, we discuss 
the HD behaviour of nuclear symmetry energy and potentials as 
well as their influences on neutron stars. Based on these
properties, we present our expectations about their influences 
on heavy-ion collisions involving neutron-rich nuclei.

\subsection{High density behaviour of nuclear symmetry energy} 
In the parabolic approximation of the ${\rm EOS}$ for neutron-rich 
matter in Eq. \ref{ieos}, the symmetry energy is\cite{wiringa88}
\begin{equation}
E_{sym}(\rho)\equiv e(\rho,1)-e(\rho,0)=\frac{5}{9}E_{kin}(\rho,0)+V_2(\rho), 
\end{equation}
where $E_{kin}(\rho,0)$ is the kinetic energy per nucleon in symmetric nuclear 
matter 
\begin{equation}\label{tf}
E_{kin}(\rho,0)=\frac{3\hbar^2}{10m}\left(\frac{3\pi^2\rho}{2}\right)^{2/3}
\end{equation}
and $V_2(\rho)$ is the deviation of the interaction energy of 
pure neutron matter from that of symmetric nuclear matter. 
The $E_{sym}(\rho)$ becomes negative if the condition\cite{kut} 
\begin{equation}
V_2(\rho)\leq -\frac{5}{9}E_{kin}(\rho,0)
\end{equation}
is reached at high densities. A negative symmetry 
energy at high densities implies that the pure neutron matter becomes 
the most stable state leading to the onset of the isospin separation 
instability ({\rm ISI}). Consequently, pure neutron domains or neutron 
bubbles surrounding isolated protons may be formed in the cores of neutron 
stars\cite{kut}. Energetic nuclear reactions with rare isotopes provide a 
unique opportunity to pin down the $E_{sym}(\rho)$ at high densities and 
study the possible existence of ISI and its consequences. 
We use the following two representatives 
of the symmetry energy as predicted by many body theories
\begin{equation}\label{esym}
E^a_{sym}(\rho)\equiv E_{sym}(\rho_0)u
\end{equation}
and
\begin{equation}
E^b_{sym}(\rho)\equiv E_{sym}(\rho_0)u\cdot\frac{u_c-u}{u_c-1},
\end{equation}
where $u\equiv\rho/\rho_0$ and $u_c=\rho_c/\rho_0$ is the reduced 
critical density at which the $E^b_{sym}(\rho)$ crosses zero and 
becomes negative at higher densities. The predicted value of $u_c$ ranges from 
about 2.7 (Hartree-Fock with the Skyrme interaction Sp\cite{mar}) 
to 9 (variational many-body approach with the
UV14+UVII interaction\cite{wiringa88}.). For our numerical calculations 
in the following we use $u_c=3$. Our conclusions in this work 
are qualitatively independent of this value. 
Thus future comparisons with experiments may rule out the existence 
of ISI below this density, but they can not exclude the occurrence
of ISI at higher densities. By using higher values for $u_c$, 
effects found here will be simply shifted to higher beam energies.
Therefore, excitation function studies in both theories and experiments 
will be necessary to determine the exact value of $u_c$ if it does exist in 
nature. The above two forms of $E_{sym}(\rho)$ 
are shown in the upper window of Fig.\ 1. By design, they both have the same 
value of $E_{sym}(\rho_0)=30$ MeV at the normal nuclear matter density $\rho_0$ 
and are very close to each other at lower densities. At high densities they 
have completely different trends reflecting the diverging predictions of 
nuclear many-body theories. 

It is necessary to mention that in a recent publication \cite{bra01}, it was
found that the symmetry enery itself is isospin-dependent at all densities.
This amounts to absorbing the higher order terms in $\delta$ into the $E_{sym}$ in the 
expansion of the EOS of neutron-rich matter in Eq.(1). Because of the charge symmetry
of nuclear interactions used the odd terms in $\delta$ should vanish. The lowest order 
correction to the parabolic approximation in Eq. (1) is the $\delta^4$ term. The magnitude
of the latter has been found extremely small in all microscopic many-body 
calculations (see, in particular those in refs. \cite{siemens70,fri81,wiringa88,bom91,sjo74,bor98}), 
except that in \cite{bra01}. The origin of the unusual prediction in the latter deserves
further investigations. In this work we stick to the parabolic approximation of the EOS
of neutron-rich matter. 

\subsection{Symmetry energy and proton fraction in neutron stars 
at $\beta$ equilibrium}
The above two forms of the symmetry energy lead to 
significantly different predictions on several properties of neutron stars. 
To a good approximation suitable for this study, a neutron star without 
neutrino trappings can be considered as a {\it npe}-matter consisting of 
nuetrons (n), protons (p) and electrons (e). Through the direct ($n\rightarrow p+e+\bar{\nu_e}$) 
and/or standard ($n+n\rightarrow n+p+e+\bar{\nu_e}$) URCA processes the $\beta$ equilibrium
\begin{equation}
n\leftrightarrow p+e^-+\bar{\nu}
\end{equation}
can be established.
The equilibrium condition requires that the respective chemical
potentials to satisfy
\begin{equation}\label{chem}
\mu_n=\mu_p+\mu_e+\mu_{\bar{\nu}},
\end{equation}
and the charge neutrality requires that 
\begin{equation}\label{dene}
\rho_p=\rho_e.
\end{equation}
Since neutrinos do not accumulate in neutron stars a few seconds after their birth
$\mu_{\bar{\nu}}=0$. Within the Fermi gas model one then finds from 
Eqs. \ref{chem} and \ref{dene} the proton fraction 
$x_{\beta}\equiv \rho_p/(\rho_n+\rho_p)$ in 
neutron stars is determined by\cite{lat91}    
\begin{equation}\label{fraction}
\hbar c(3\pi^2\rho x_{\beta})^{1/3}=4E_{\rm sym}(\rho)(1-2x_{\beta}).
\end{equation}   
The equilibrium proton fraction is therefore entirely determined by 
the $E_{sym}(\rho)$. The values of $x_{\beta}$ corresponding to 
the two chosen forms of symmetry energy are shown in the lower window of Fig.\ 1.
With the $E^{b}_{sym}(\rho)$, the $x_{\beta}$ is zero for $\rho/\rho_0\geq 3$, 
indicating that a pure neutron matter could become most stable, 
leading to the isospin separation instability in HD neutron-rich matter. 
The value of $x_{\beta}$ influences many other properties of neutrons stars, 
such as the cooling rate of protoneutron stars\cite{lat91}. The fast 
cooling mechanism through the direct URCA process can happen if $x_{\beta}$ is 
larger than about $1/9$. It is seen that with the $E^{a}_{sym}(\rho)$, 
the neutron star becomes so proton-rich that the 
fast cooling can happen at densities higher than about $2.3\rho_0$. On the
contrary, it is impossible for this process to happen with the 
$E^{b}_{sym}(\rho)$.

\subsection{Equation of state of neutron-rich matter}
With the two forms of the symmetry energy $E^a_{sym}(\rho)$ and 
$E^b_{sym}(\rho)$, the corresponding {\rm EOS} for neutron-rich matter 
is rather different. To illustrate this point, 
the simplest, momentum-independent parameterization
\begin{equation}
e(\rho,0)= \frac{a}{2}u+\frac{b}{1+\sigma}u^{\sigma}+\frac{3}{5}e_F^0u^{2/3}
\end{equation}
is used as the isoscalar part of the {\rm EOS}, 
where $u\equiv \rho/\rho_0$ is the reduced density and $e_F^0=36$ MeV is the Fermi 
energy. The parameters $a=-358.1$ MeV, $b=304.8$ MeV and $\sigma=7/6$ 
are determined by saturation properties and a compressibility 
$K_{\infty}=201$ MeV of isospin symmetric nuclear matter. 
Our conclusions in this work are qualitatively independent of the particular 
form of $e(\rho,0)$. The use of a stiffer {\rm EOS}, such as $K_{\infty}=380$ MeV,
will reduce the compression by about 15\% in the energy range studied in this work. This
will then simply shift the observed isospin effects to slightly higher beam energies.
Shown in Fig.\ 2 are the ${\rm EOS}$ of isospin asymmetric 
nuclear matter with the $E^a_{sym}(\rho)$ (upper window) and  $E^b_{sym}(\rho)$ 
(lower window), respectively. With the linearly increasing $E^a_{sym}(\rho)$, 
the ${\rm EOS}$ becomes stiffer with the increasing $\delta$. The isospin 
symmetric nuclear matter remains to be the ground state at all densities. 
This is in stark contrast to the situation using the $E^b_{sym}(\rho)$. 
The ${\rm EOS}$ obtained with the latter is softened instead of being
stiffened by the increasing isospin asymmetry $\delta$ at densities 
higher than $3\rho_0$. At these high densities, in agreement with the information extracted
from Fig.\ 1, the pure neutron matter becomes the most stable state whereas 
the isospin symmetric nuclear matter becomes the most unstable state of HD 
nuclear matter.

\subsection{Nuclear symmetry potentials and their softest lines}
In accordance with the {\rm EOS} discussed above, the isoscalar potential is
\begin{equation}
v_0(\rho)=au+bu^{\sigma}
\end{equation}
and the symmetry potentials can be obtained from 
\begin{equation}\label{vasy} 
v_{\rm asy}^{q}=\frac{\partial W_{asy}}{\partial \rho_{q}}
\end{equation}
where {\rm q=neutrons} and {\rm q=protons}, respectively. 
The symmetry potential energy density $W_{asy}$ is   
\begin{equation} 
W_{asy}=V_2\rho\delta^{2}=\left[E_{sym}(\rho)-c_1u^{2/3}\right]\rho\delta^{2},
\end{equation} 
where $c_1=3/5e_F^0(2^{2/3}-1)\approx 12.7$ MeV.
With the $E^{a}_{sym}(\rho)$ one obtains 
\begin{equation}
v_{\rm asy}^{a}=\pm 2(E_{sym}(\rho_0)-c_1)u\delta,
\end{equation}
where ``+" and ``-" signs are for neutrons and protons, respectively. 
With the $E^{b}_{sym}(\rho)$, the symmetry potential is 
\begin{equation}
v_{\rm asy}^{b}=\pm 2(c_2u_cu-c_2u^2-c_1u^{2/3})\delta
+(1/3c_1u^{2/3}-c_2u^2)\delta^2,
\end{equation}
where $c_2=E_{sym}(\rho_0)/(u_c-1)\approx 15$ MeV. 
At low densities $v^a_{asy}$ and $v^b_{asy}$ are very close since
\begin{eqnarray}
\lim_{u\leq 1}v_{asy}^b&\approx& \pm 2\left[c_2(u_c-u)-c_1u^{-1/3}\right]u\delta
\\\nonumber
&\approx& \pm 2\left[c_2(u_c-1)-c_1\right]u\delta\\\nonumber
&=&\pm 2(E_{sym}(\rho_0)-c_1)u\delta=v_{asy}^a.
\end{eqnarray} 
This is so because of our choice of the two forms of the symmetry energy 
with the desire to study its HD behaviour. This feature is further illustrated in
Fig.\ 3 where the above two symmetry potentials are shown as a function of density
for $\delta=0.2$ and $u_c=3$. As one expects the symmetry potentials are rather
similar at $\rho\leq \rho_0$ but very different at higher densities for both neutrons
and protons. The different HD behaviour is expected to cause neutrons and protons to 
behave differently during the dynamical evolution of heavy-ion collisions.

Our main purpose in the following will be searching 
for signs of the different symmetry potentials in experimental 
observables of high energy heavy-ion collisions by comparing results
for neutrons and protons and using different symmetry potentials. Of course, 
what is more important for the reaction dynamics is the force, or 
the slope of potentials.
With the $E^{a}_{sym}(\rho)$ 
the repulsive (attractive) symmetry 
potential for neutrons (protons) increases linearly. 
With the $E^{b}_{sym}(\rho)$, however, the symmetry potential 
for neutrons (protons) has a broad maximum (minimum) around 
$\rho=1.2\rho_0$ ($\rho=1.4\rho_0$). Thus for both neutrons and protons 
the forces due to the symmetry potentials are zero around the extreme of their
respective symmetry potentials. This property then leads to a {\it softest point} 
in the resultant {\rm EOS} at a given $\delta$ for isospin asymmetric 
nuclear matter. Generally, for a varying $\delta$ there is a {\it softest line} 
for both neutrons and protons. These lines are determined by the condition 
\begin{equation}
(\partial v^b_{asy}/\partial \rho)_{\delta}=0.
\end{equation}
With the $E^b_{sym}(\rho)$ these lines are along 
\begin{equation}
\delta_{softest}=\pm \frac{6c_1+18c_2u^{4/3}-9c_2u_cu^{1/3}}
{c_1-9c_2u^{4/3}},
\end{equation}
where the $\pm$ sign is for neutrons/protons.
These {\it softest lines} in neutron-rich matter are shown in 
Fig.\ 4.  In heavy-ion collisions with a proper combination of 
the beam energy and isospin asymmetry, the system can cross these softest lines.
The crossing points may manifest themselves as local minima in the
excitation functions of collective flow for both neutrons and protons. 
This expectation is analogous to that for the local minimum in the 
excitation function of transverse flow due to the Quark-Gluon-Plasma 
induced softening of the {\rm EOS} in ultra-relativistic 
heavy-ion collisions\cite{soft}. The Coulomb potential may 
shift the relative values and slopes of proton potentials 
depending on the dynamical evolution of the charge distributions 
during the reaction. It is just difficult to predict quantitatively where
the minima may appear based on the discussions of the symmetry potentials.
Fortunately, transport models provide a reliable tool to make more quantitative 
predictions possible.

\section{Isospin asymmetry of dense nuclear matter formed in high energy 
heavy-ion collisions} 
High energy heavy-ion collisions provide the only terrestrial situation 
where the HD neutron-rich matter can be formed. Moreover, fast 
radioactive heavy-ion beams to be available at the planned Rare Isotope 
Accelerator (RIA) in the United States can make the HD nuclear 
matter even more neutron-rich\cite{ria}. The HD 
behaviour of $E_{sym}(\rho)$ affects properties of the HD 
nuclear matter formed in high energy heavy-ion collisions.
Moreover, interesting precursors of the ISI might be 
observed in collisions with energetic neutron-rich nuclei.

\subsection{Summary of an isospin-dependent hadronic transport model} 
We investigate the effects and phenomena mentioned above within an 
isospin-dependent hadronic transport model for an interacting system 
of nucleons, Delta resonances and pions\cite{ibuu2,ibuu1,ibuu3}. 
Evolutions of the phase-space distribution functions of nucleons, 
Delta resonances and pions with their explicit isospin degrees 
of freedom are solved numerically by using of the test-particle 
approach\cite{wong,bet}. Isospin-dependent total and differential cross 
sections among all particles are taken either from the elementary particle 
scattering data or obtained by using the detailed balance\cite{dani,bali}. 
Explicitly isospin-dependent Pauli blockings for fermions are also employed. 
For a review of the model, we refer the reader to refs.\cite{lkb98,li01}. 

At beam energies above the pion production threshold, baryon resonances
play an import role in the reaction dynamics. In this study we limit ourselves
to beam energies less than 2 GeV/nucleon. It is worth mentioning that 
for medium sized rare isotopes, such as $^{32}Na$, 
high energy beams up to about 950 MeV/nucleon are currently available 
at GSI and for heavy rare isotopes, such as $^{132}Sn$, 
beams up to 400 MeV/nucleon will be available at the future RIA. 
In this energy range, the most important baryon resonance is the $\Delta (1232)$. 
Shown in Fig.\ 5 are two examples of the multiplicities of $\Delta (1232)$ 
resonances and pions. The maximum population of the $\Delta (1232)$ is 
about 2\% and 20\% in the
$^{132}Sn+^{124}Sn$ reaction at an impact parameter of 1 fm and a beam energy 
of 400 and 2000 MeV/nucleon, respectively. The mean field potentials of pions and 
baryon resonances in nuclear matter are still largely unknown. We thus make 
here a minimum assumption that the isoscalar part of the $\Delta$ potential is the same 
as that for nucleons. To be consistent with the modeling of the isovector
potential for nucleons, we assume that the isovector potential for 
$\Delta$ resonances is an average of that for neutrons and protons. 
The weighting factor depending on the charge state of the resonance 
is the square of the Clebsch-Gordon coefficients for isospin coupling in the processes 
$\Delta\leftrightarrow \pi N$. Thus, we have
\begin{eqnarray}
v_{asy}(\Delta^-)&=&v_{asy}(n),\\\
v_{asy}(\Delta^0)&=&\frac{2}{3}v_{asy}(n)+\frac{1}{3}v_{asy}(p),\\\ 
v_{asy}(\Delta^+)&=&\frac{1}{3}v_{asy}(n)+\frac{2}{3}v_{asy}(p),\\\ 
v_{asy}(\Delta^{++})&=&v_{asy}(p).
\end{eqnarray} 
Similarly, we define the effective isospin asymmetry $\delta_{like}$ 
for excited baryonic matter as
\begin{equation}
\delta_{like}\equiv \frac{(\rho_n)_{like}-(\rho_p)_{like}}
{(\rho_n)_{like}+(\rho_p)_{like}},
\end{equation}
where 
\begin{eqnarray}
(\rho_n)_{like}&=&\rho_n+\frac{2}{3}\rho_{\Delta^0}+\frac{1}{3}\rho_{\Delta^+}
+\rho_{\Delta^-},\\\
(\rho_p)_{like}&=&\rho_p+\frac{2}{3}\rho_{\Delta^+}+\frac{1}{3}\rho_{\Delta^0}
+\rho_{\Delta^{++}}.
\end{eqnarray}
It is evident that the $\delta_{like}$ reduces naturally to $\delta$ as the
beam energy becomes smaller than the pion production threshold.

\subsection{Isospin asymmetry of dense matter formed in high energy 
heavy-ion collisions} 
Shown in Fig.\ 6 is the evolution of the central baryon density for the reaction 
of $^{132}Sn+^{124}Sn$ at an impact parameter of 1 fm and beam energies of 
200, 400, 1000 and 2000 MeV/nucleon, respectively. In this energy range a 
compression of about 1.7 to 3.7 times the normal nuclear matter density is reached. 
This very reaction at beam energies upto about 400 MeV/nucleon will be available at the 
RIA facility\cite{ria}. 
One notices that the $E^{b}_{sym}(\rho)$ leads to a slightly higher compression 
of about 10-15\%. This is due to the relatively softened nuclear ${\rm EOS}$ 
with the $E_{sym}^b$ than that with the $E_{sym}^a$.
   
Is the compressed hadronic matter neutron-rich or -poor compared to the initial
reaction system? The answer to this question has important implications to several
critical questions in nuclear astrophysics. To answer this question we show 
in Fig.\ 7 the correlation between the baryon density and the isospin asymmetry
$\delta_{like}$ over the entire reaction volume at the time of about the maximum 
compression in the central $^{132}Sn+^{124}Sn$ reactions with E/A=400 (middle window) 
and 2000 (bottom window) MeV/nucleon, respectively. The whole reaction volume is 
divided into cubic cells of 1 {\rm fm}$^3$ in size. Each point in the scatter plot 
represents one such cell. For a comparison, the correlation in the initial state 
of the reaction is shown in the upper window. The initial state for our transport model 
calculations is generated by using neutron and proton density profiles predicted 
by the Relativistic Mean Field model\cite{serot,ren}. The average $\delta$ of the initial
reaction system is 0.22, the low density tail extending to the far more
neutron-richer side is due to the neutron skin in both the target and 
projectile nuclei. Thus a significant number of cells with densities close to 
$\rho/\rho_0=1$ have isospin asymmetries less than 0.22. 
How this initial $\rho-\delta$ correlation evolves in a heavy-ion collision 
depends sensitively on the HD behaviour of nuclear 
symmetry energy. With the $E^a_{sym}(\rho)$ the trend of continuous neutron 
distillation from higher density regions to 
lower ones persists at all energies. The high density region around 
$\rho=2\rho_0$ is about twice more isospin symmetric than that with 
the $E^b_{sym}(\rho)$ at 400 MeV/nucleon.  Most interestingly, with 
the $E^b_{sym}(\rho)$ the onset of ISI is clearly seen as indicated by 
the right turn of $\delta_{like}$ to higher values at $\rho\geq 3\rho_0$ in 
the reaction at $E_{beam}/A=2$ GeV/nucleon. 

The reason behind the observed neutron-distillation from the HD 
baryonic matter can be easily understood from the density dependence of 
symmetry energy shown in the upper window of Fig.\ 1.  For two parts (1 and 2) 
of isospin asymmetric nuclear matter to be in chemical equilibrium the condition 
\begin{equation}\label{mig}
E_{sym}(\rho_1)\delta_1=E_{sym}(\rho_2)\delta_2
\end{equation}
must be satisfied\cite{muller,liko,baran,shi}. Thus, it is energetically 
favorable in a dynamical process to have a migration of nucleons 
with its direction determined by Eq. \ref{mig} according to the 
density dependence of the $E_{sym}(\rho)$. 
With the $E^a_{sym}(\rho)$ which increases linearly with $\rho$, 
there is a continuous migration of neutrons (protons) 
from higher (lower) to lower (higher) densities. Whereas with the 
$E^b_{sym}(\rho)$ which peaks at $u=1/2u_c$, it is most energetically 
favorable to deplete (concentrate) 
all neutrons (protons) from (to) the peak. Moreover, at 
$u\geq u_c$ the ISI sets in. It leads to an unlimited trapping of neutrons above
the critical density $u_c=3$. 

To be more clear about the roles of the density dependent symmetry energy 
we investigate in Fig. 8 the average (over all cells of the same 
density) $\delta_{like}$ as a function of density for the central $^{132}Sn+^{124}Sn$ 
reactions at 400 and 2000 MeV/nucleon. The overall rise of $\delta$ at low 
densities is mainly due to the neutron skins of the colliding nuclei and the 
distillated neutrons. Effects due to the different symmetry energies are more clearly 
revealed especially at high densities.
For a comparison and study implications of our results 
to nuclear astrophysics, the $\rho-\delta$ correlation in neutron 
stars at $\beta$ equilibrium is shown in the insert of Fig.\ 8. 
The isospin asymmetry at $\beta$ equilibrium $\delta_{\beta}=1-2x_{\beta}$ is 
completely determined by the $E_{sym}(\rho)$. With the $E^{b}_{sym}(\rho)$, 
the $\delta_{\beta}$ is $1$ for $\rho/\rho_0\geq 3$, 
indicating that the neutron star has become a pure neutron matter at 
these high densities. On the contrary, with the $E^{a}_{sym}(\rho)$, 
the neutron star becomes more proton-rich as the density increases.
An astonishing similarity is seen in the resultant 
$\delta-\rho$ correlations for the neutron star and the heavy-ion collision. 
In both cases, the symmetry energy $E^b_{sym}(\rho)$ makes the HD nuclear 
matter more neutron-rich than the $E^a_{sym}(\rho)$ and the effect 
grows with the increasing density. Of course, this is no surprise 
since the same underlying nuclear ${\rm EOS}$ is at work in both cases. 
The decreasing $E^b_{sym}(\rho)$ above $\frac{1}{2}u_c=1.5\rho_0$ makes 
it more energetically favorable to have the denser region more neutron-rich. 

The above analyses of the  main features are at the instants of about the 
maximum compressions. How do they evolve in time?
To answer this question the average $n/p$ ratio of the HD region 
with $\rho/\rho_0\geq 1$ is shown as a function of time and beam energy 
in Fig.\ 9. The effect on $(n/p)_{\rho/\rho_0\geq 1}$ 
due to the different $E_{sym}(\rho)$ is seen to grow with the reaction
time until the expansion has lead the system to densities below $\rho_0$, 
especially at higher beam energies. Although the compression starts at 
about the same time, the expansion starts on a faster time scale at higher 
beam energies as one expects. Again, it is seen that whether the HD 
region is neutron-rich or -poor depends critically on the HD behaviour 
of nuclear symmetry energy.  

\subsection{Pion production effects on n/p ratio of dense matter}
During heavy-ion collisions there is a net 
conversion of neutrons to protons due to the more abundant 
production of $\pi^-$'s than $\pi^+$'s. It is 
therefore also interesting to explore effects of pion production on the
n/p ratio of dense matter. This study is also necessary to be certain about
the effects of the HD nuclear symmetry energy. We thus compare in Fig. 10 
the HD n/p ratio with and without including the pion production channel in the 
central $^{132}Sn+^{124}Sn$ reaction at a beam energy of 2 GeV/nucleon. The beam
energy is chosen to illustrate the maximum effects of pion production since the
latter decreases at lower beam energies. We artificially turned off the 
pion production channel, i.e., $N+N\rightarrow N+\Delta$, and attribute 
its cross section to elastic nucleon-nucleon scatterings such that 
the total reaction cross section is kept a constant. The central densities 
are shown in the upper rows of Fig. 10. It is seen that the overall 
compression is slightly reduced without including the pion production. 
This is because the latter 
has the role of enhancing the stopping power. It is interesting to note that 
the effect (difference between the curves a and b) on compression due to 
the different symmetry energy is about the same with or without the pion 
production. The lower two rows demonstrate effects of pion production on 
the HD n/p ratio. By comparing the left and right windows of the lower two rows, 
it is seen that the pion production reduces the HD n/p ratios by about 15\%, 
while the sensitivity to the symmetry energy remains almost the same. 
This finding also indicates that the pion production might be used as a 
perturbative probe of the n/p ratio of HD nuclear matter.   

\section{Probing the HD behaviour of nuclear symmetry energy with high energy
heavy-ion collisions}
How to probe experimentally the HD behaviour of $E_{sym}(\rho)$ 
in high energy heavy-ion collisions? In this section we study four complementary
approaches. These are the $\pi^-/\pi^+$ ratio, nuclear transverse 
collective flow and its excitation function as well as the neutron-proton 
differential flow.

\subsection{Pion probe}
As we have shown in the previous section, that the pion production only affects  
perturbatively the $n/p$ ratio of HD nuclear matter. On the other hand, 
at beam energies below about 2 GeV/nucleon, pions are mostly produced through the decay 
of $\Delta(1232)$ resonances. The primordial $\pi^-/\pi^+$ ratio is 
approximately quadratic in n/p according to the branching ratios of single pion 
production via $\Delta$ resonances in nucleon-nucleon collisions
\begin{equation}
\pi^-/\pi^+=\frac{5n^2+np}{5p^2+np}\approx (n/p)^2.
\end{equation}
Thus the $\pi^-/\pi^+$ ratio is expected to be a sensitive probe of the 
n/p ratio of HD nuclear matter. 
Pion reabsorptions and rescatterings ($\pi+N\rightarrow \Delta$ and 
$N+\Delta\rightarrow N+N$) are expected to 
complicate the above relationship. Nevertheless, 
very high sensitivity to the $n/p$ ratio is retained in 
the final $\pi^-/\pi^+$ ratio as indicated in the experimental 
data of high energy heavy-ion collisions\cite{stock}. 
Shown in Fig.\ 11 are the $(\pi^-/\pi^+)_{like}$ 
ratio
\begin{equation}
(\pi^-/\pi^+)_{like}\equiv \frac{\pi^-+\Delta^-+\frac{1}{3}\Delta^0}
{\pi^++\Delta^{++}+\frac{1}{3}\Delta^+}
\end{equation} 
as a function of time for the central $^{132}Sn+^{124}Sn$ reaction at 
beam energies from 200 to 2000 MeV/nucleon. This ratio naturally 
becomes the final $\pi^-/\pi^+$ ratio at the freeze-out 
when the reaction time $t$ is much longer 
than the lifetime of the delta resonance $\tau_{\Delta}$.
The $(\pi^-/\pi^+)_{like}$ ratio is rather high in the early 
stage of the reaction because of the large numbers of neutron-neutron 
scatterings near the surfaces where the neutron skins of the colliding nuclei overlap. 
By comparing Fig.\ 9 and Fig.\ 11, it is seen that a variation of about 
30\% in the $(n/p)_{\rho/\rho_0\geq 1}$ ratio due to the different $E_{sym}(\rho)$ 
results in about 15\% change in the final $\pi^-/\pi^+$ ratio.
It has thus an appreciable response factor of 
about 0.5 to the variation of HD n/p ratio and is approximately 
beam energy independent. Therefore, one can conclude that 
the $(\pi^-/\pi^+)_{like}$ ratio is a direct probe of the HD n/p ratio, and thus
an indirect probe of the HD behaviour of nuclear symmetry energy.

\subsection{Transverse collective flow and its excitation function}
Next we investigate the transverse collective flow as a probe of the HD 
behaviour of nuclear symmetry energy. 
We perform the standard analysis of the average transverse momentum 
in the reaction plane\cite{pawel85} 
\begin{equation}
< p_x/N >(y)=\frac{1}{N(y)}\sum_{i=1}^{N(y)} p_{ix}(y),
\end{equation}
where $N(y)$ is the number of particles at rapidity $y$ and $p_{ix}$ is $i^{th}$ 
particle's transverse momentum in the reaction plane. Shown in Fig.\ 12 are 
two typical examples of the transverse collective flow analysis for neutrons 
and protons for the mid-central $^{132}Sn+^{124}Sn$ 
reactions at 400 MeV/nucleon. Appreciable effects of the nuclear symmetry 
energy is clearly seen, especially for protons. 

We found that the symmetry energy effects on the transverse flow 
vary with beam energy and are different for 
neutrons and protons. To better characterize the effects 
we study in Fig.\ 13 the excitation function of flow parameter. The latter is 
defined as the slope of the transverse momentum distribution at mid-rapidity, i.e. 
\begin{equation}
F\equiv \left(\frac{d<p_x/N>}{dY_{cm}/y_{beam}}\right)_{y_{cm}=0}.
\end{equation}
We concentrate on analyzing differences between the collective flows of 
neutrons and protons as well as those caused by using the different symmetry
energies. These differences can be mainly attributed to the different 
symmetry potentials for neutrons and protons since the isoscalar potential has
no direct contribution to these differences to the first order of approximation. 
On top of the generally growing flow parameter there are 
distinct structures closely associated with the symmetry energy used.
In particular, a local minimum appears in the excitation functions for 
both neutrons and protons around $E_{beam}/A=500$ MeV with the $E_{sym}^b$. 
The dip is more obvious for protons than neutrons. 
These minima are direct signals of the softening of the underlying {\rm EOS}
as we expected from studying the softest lines of the symmetry potentials. 
With the $E_{sym}^a$ there is also a local minimum in the excitation function for 
protons around $E_{beam}/A=300$ MeV. This minimum is a result of the 
interplay between repulsive hadronic scatterings and the attractive symmetry 
potential for protons (the solid line in the bottom window of Fig. 3). To 
be more quantitative, the average number of successful 
hadronic scatterings per nucleon and the maximum baryonic 
compression are shown as a function of beam energy in Fig.\ 14. 
Because the $E_{sym}^b$ softens the {\rm EOS} it thus leads to an 
about 3\% higher compression and more scatterings compared to the 
case with the $E_{sym}^a$. The small increase in compression has very
little effect on the scalar potential. Thus the observed structures
can not be due to the isoscalar potentials. Below the pion production 
threshold in nucleon-nucleon collisions at about 296 MeV, the Pauli 
blocking is still strong and thus the collision numbers increase very 
little with beam energy. 
Whereas above the pion production threshold, because of the opening of the 
inelastic channel via $N+N\rightarrow N+\Delta$ 
and its resonance nature, the collision numbers grow quickly. 
With the $E_{sym}^a$, for protons the symmetry potential causes negative flow 
while the collisions result in positive flow.  Around 300 MeV/nucleon
the collision numbers are still small, effects of the symmetry potential 
are thus relatively important in the interplay which leads to the dip near
the pion production threshold. At higher beam energies, however, the flow 
increases because of the dominating role of the collisions and their fast growth with 
beam energy. While for neutrons with the $E_{sym}^a$, both the collisions 
and the repulsive symmetry potential (the solid line in the upper window of Fig.\ 3) 
enhance the transverse flow. Thus the flow parameter for neutrons increases almost 
monotonously with beam energy. Therefore, the structures observed above in the 
excitation functions of the nuclear transverse flow parameter are unique qualitative
signatures of the density dependence of symmetry energy.  
  
\subsection{Neutron-proton differential flow probe}
The neutron-proton differential collective flow is measured
by\cite{li00} 
\begin{equation}
F_{np}(y)\equiv\frac{1}{N(y)}\sum_{i=1}^{N(y)}p_{x_i}\tau_i,
\end{equation}
where $N(y)$ is the total number of free nucleons at the rapidity $y$, 
$p_{x_i}$ is the transverse momentum of particle $i$ in the reaction 
plane, and $\tau_i$ is $+1$ and $-1$ for neutrons and protons, respectively.
The free nucleons are identified as those having local nucleon densities less
than $1/8\rho_0$. The $F_{np}(y)$ combines constructively the in-plane transverse 
momenta generated by the isovector potentials while reducing 
significantly influences of the isoscalar potentials of both neutrons 
and protons. Thus, it can reveal more directly the HD behaviour 
of $E_{sym}(\rho)$ in high energy heavy-ion collisions. Two typical 
results for mid-central $^{132}Sn+^{124}Sn$ reactions at 400 MeV/nucleon 
and 1000 MeV/nucleon, respectively, are shown in Fig.\ 15. A clear 
signature of the HD behaviour of $E_{sym}(\rho)$ 
appears at both forward and backward rapidities. To characterize 
the effect, the slope $dF_{np}/d(y_{cm}/y_{beam})$ 
at mid-rapidity is shown as a function of beam energy in Fig.\ 16. 
A striking difference of about a factor of 2 exists in the 
reactions at $E_{beam}\geq 200$ MeV/nucleon. This large effect 
can be very easily observed by using available detectors at several heavy-ion 
facilities in the world. However, the interesting structures observed in the 
excitation function of the transverse flow parameter $F$ is largely being 
smeared out because the different rapidity distributions of neutrons and protons
also enters in the calculation of the neutron-proton differential flow. 
Compared with the $\pi^-/\pi^+$ ratio, both the transverse flow and the 
neutron-proton differential collective flow are more directly affected by and 
are thus also more sensitive probes of the HD behaviour of $E_{sym}(\rho)$. 

\section{Summary}
In summary, the HD behaviour of nuclear symmetry energy has been 
puzzling physicists for decades. In this work, high energy 
heavy-ion collisions are proposed as a novel means 
to solve this longstanding problem. Within an isospin dependent hadronic 
transport model using two representative density dependent symmetry energy 
functions predicted by many body theories, it is shown 
that the isospin asymmetry of HD nuclear matter formed in high energy heavy-ion 
collisions is uniquely determined by the HD behaviour of $E_{sym}(\rho)$. 
It is demonstrated that the isospin separation instability indeed can happen
in high energy heavy-ion collisions. We explored several promising experimental 
probes for the HD symmetry energy. Among them, four quantitatively sensitive
observables, the $\pi^-/\pi^+$ ratio, nuclear transverse collective flow and
its excitation function as well as the neutron-proton differential collective 
flow are studied. A precursor of the possible isospin separation instability 
in dense neutron-rich matter is predicted to appear as the local minima in the 
excitation functions of the flow parameter for both neutrons and protons above 
the pion production threshold. This precursor can be used as a unique signature 
of the isospin dependence of the nuclear {\rm EOS}. This {\it qualitative} signature is 
advantageous over the other {\it quantitative} observables because the latter 
are often affected by several ingredients of the reaction dynamics and 
experimental uncertainties. It should be noted that the present study is based
on a momentum-independent transport model. The momentum-dependent potential is known
to be important for a reliable description of global momentum distributions. 
Nevertheless, we expect that all qualitative features, such as the existence of the minimum 
in the excitation function of transerse flow, are not affected by the momentum-dependence. 
However, the quantitative results, such as the exact location of the flow minimum, 
are affected by the momentum-dependence of the nuclear potential. A more refined study 
using a momentum-dependent isoscalar potential is underway and the results will be reported 
elsewhere. It should also be mentioned that the fragmentation mechanism of isospin 
asymmetric nuclear matter is still not clear and remains a hot topic of current 
studies\cite{li02a}. The different fragmentation mechanisms might affect the neutron/proton 
ratio of the observed free nucleons, and thus the neutron-proton differential flow. It would 
thus be useful to measure also observables associated with fragments together with 
the single particle observables examined in this study. It still remains a theoretical 
challenge to develop a fully quantum transport theory that deals with both particle 
production and cluster formation in a numerically tractable manner. The present work 
shows the need and promise of working in this direction. Experimental measurements of the 
proposed observables will provide the first terrestrial data to constrain stringently 
the HD behaviour of nuclear symmetry energy. Future comparisons between the experimental 
data and model calculations will allow us to extract crucial information about the {\rm EOS} 
of dense neutron-rich matter. 

This work was supported in part by the National Science Foundation Grant 
No. PHY-0088934 and Arkansas Science and Technology Authority Grant No. 00-B-14.

\newpage

\begin{figure}[htp] 
\vspace{3.5cm}
\centering \epsfig{file=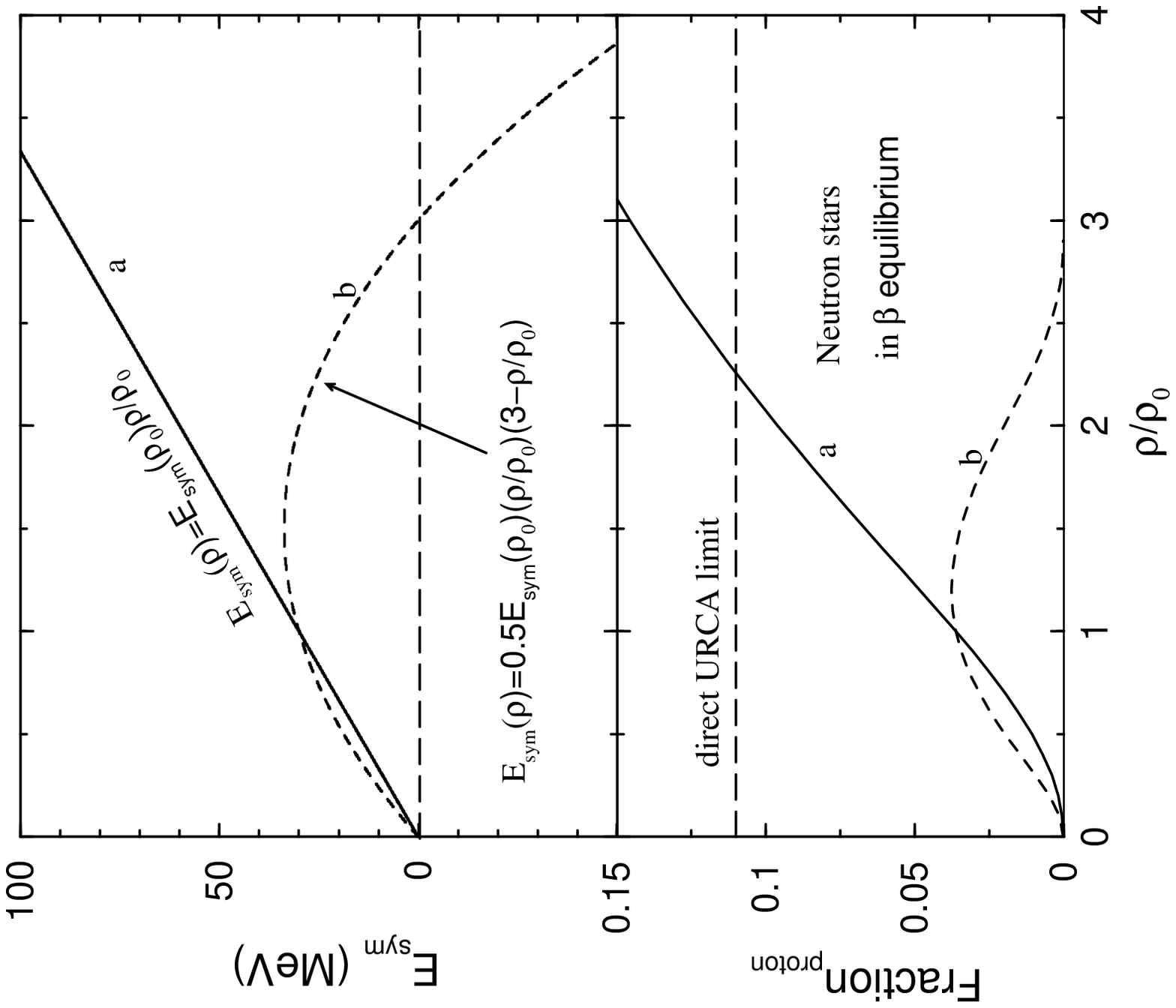,width=14cm,height=10cm,angle=-90} 
\vspace{1.cm}
\caption{Upper window: Two representatives of the nuclear symmetry energy
as a function of density. Lower window: the corresponding proton fractions in 
neutron stars at $\beta$ equilibrium.} 
\label{fig1}
\end{figure}

\begin{figure}[htp] 
\vspace{3.5cm}
\centering \epsfig{file=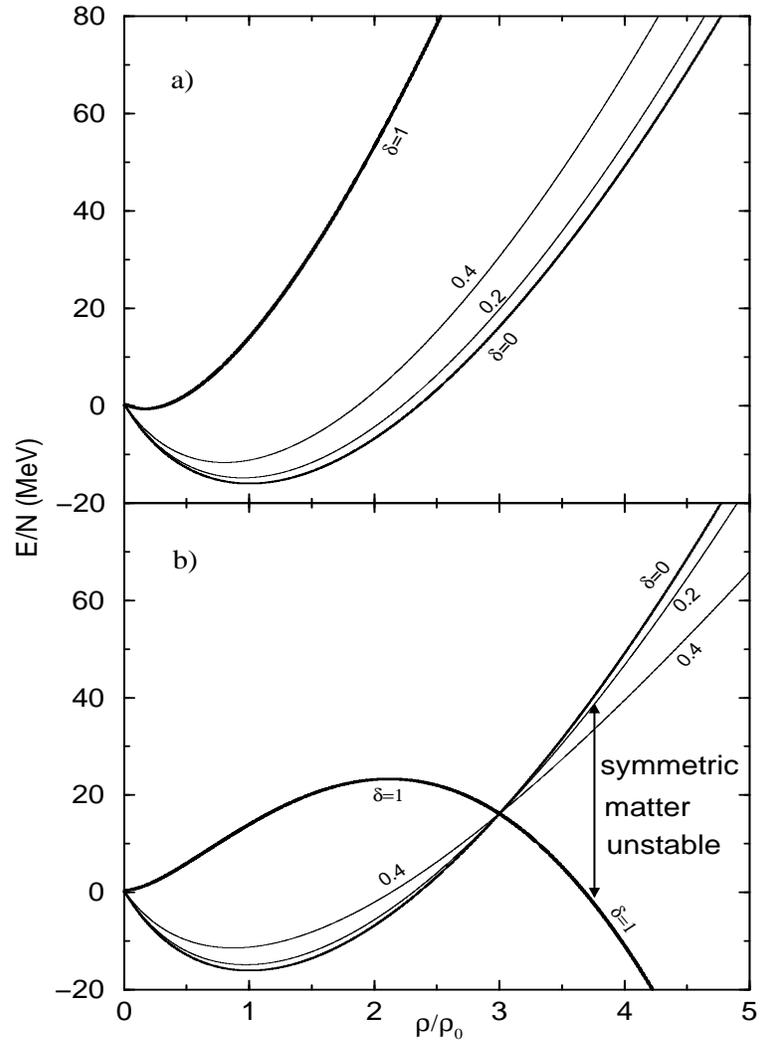,width=14cm,height=10cm,angle=-90} 
\vspace{1.cm}
\caption{The upper (lower) window is the equation of state of 
isospin-asymmetric nuclear matter using the 
nuclear symmetry energy parameterization $E^a_{sym}$ ($E^b_{sym})$.} 
\label{fig2}
\end{figure}

\begin{figure}[htp] 
\vspace{3.5cm}
\centering \epsfig{file=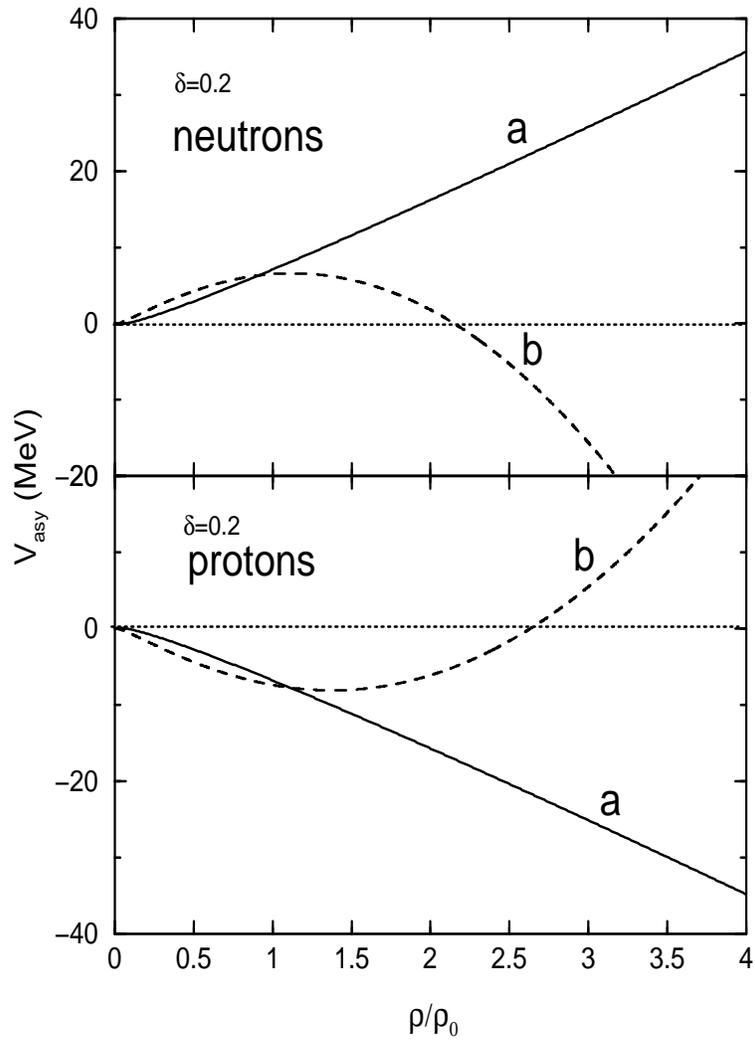,width=14cm,height=10cm,angle=-90} 
\vspace{1.cm}
\caption{Symmetry potentials for neutrons (upper window) and 
protons (lower window) using the nuclear symmetry energy 
$E^a_{sym}$ and $E^b_{sym}$, respectively.} 
\label{fig3}
\end{figure}

\begin{figure}[htp] 
\vspace{3.5cm}
\centering \epsfig{file=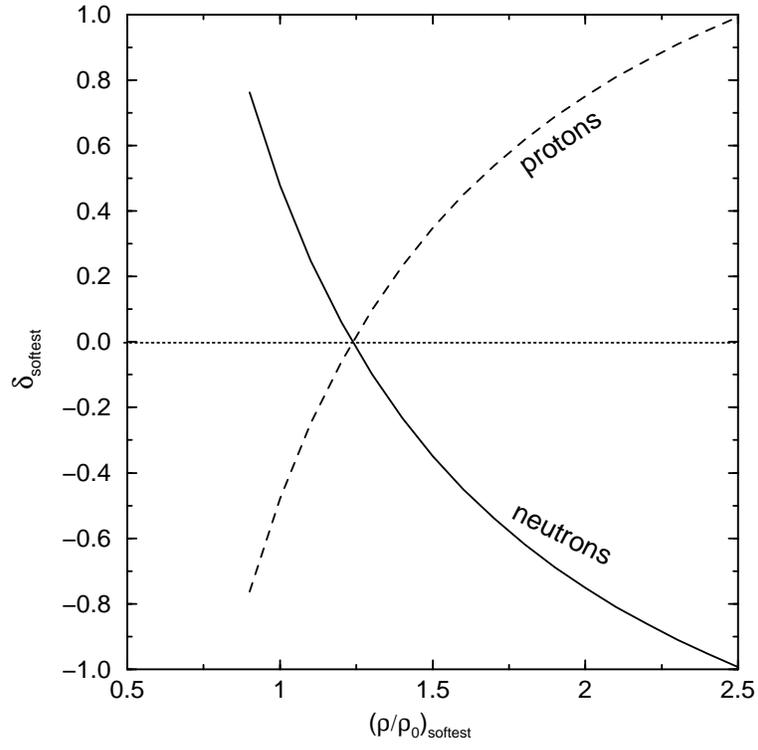,width=10cm,height=10cm,angle=-90} 
\vspace{1.cm}
\caption{The softest lines for neutrons and protons using 
the symmetry potential corresponding to the $E^b_{sym}$.} 
\label{fig4}
\end{figure}

\begin{figure}[htp] 
\vspace{3.5cm}
\centering \epsfig{file=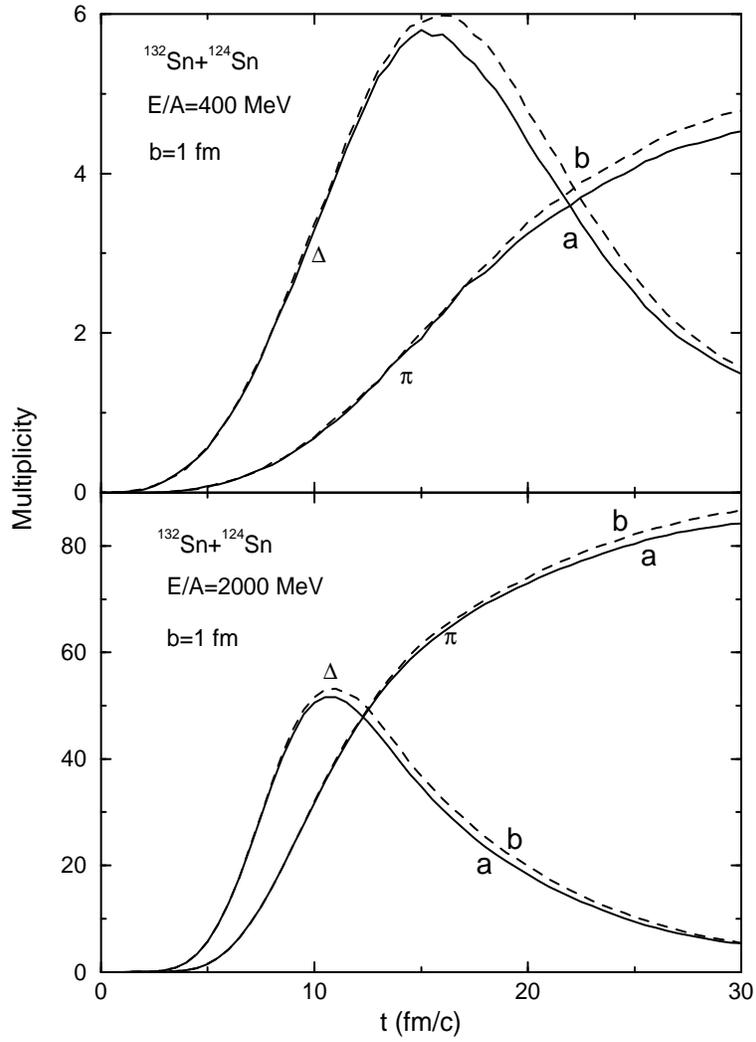,width=14cm,height=10cm,angle=-90} 
\vspace{1.cm}
\caption{Multiplicities of Delta resonances and pions in the central 
reaction of $^{132}Sn+^{124}Sn$ at a beam energy of 400 MeV/nucleon (upper window) and
2 GeV/nucleon (lower window).} 
\label{fig5}
\end{figure}

\begin{figure}[htp] 
\vspace{3.5cm}
\centering \epsfig{file=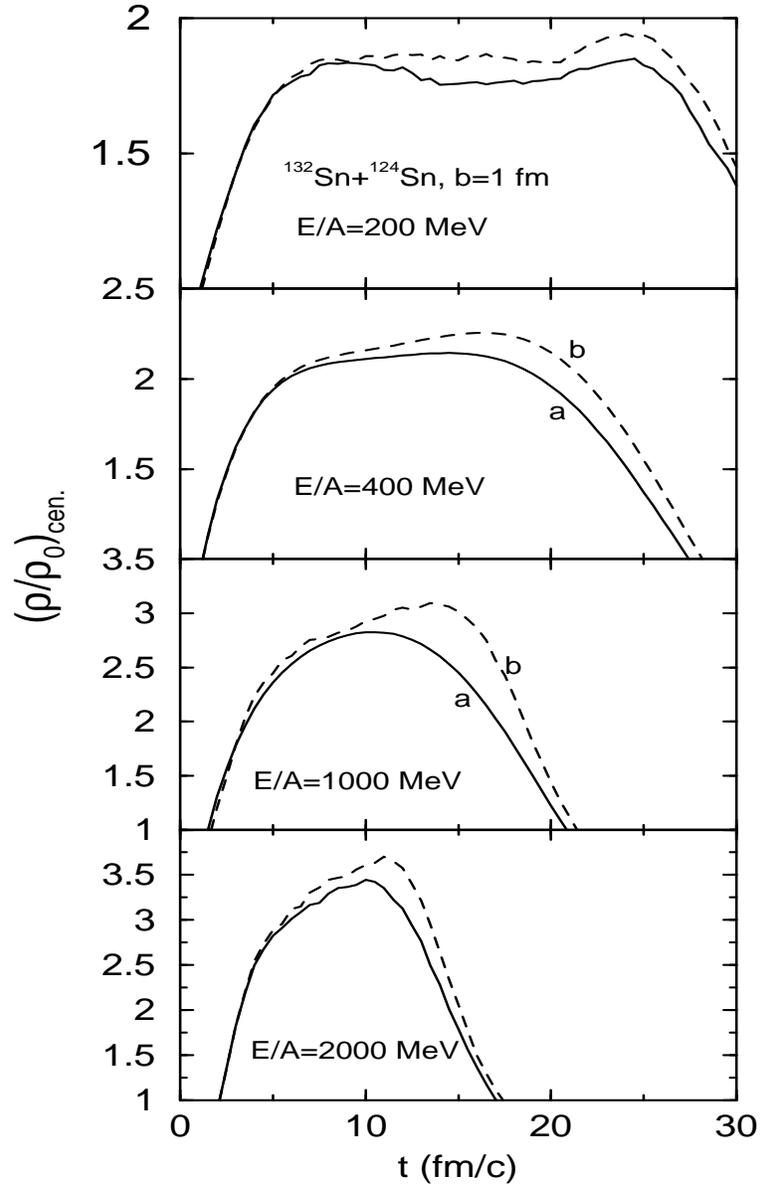,width=16cm,height=10cm,angle=-90} 
\vspace{1.cm}
\caption{Evolution of the central baryon density in the central 
reaction of $^{132}Sn+^{124}Sn$ at a beam energy from 200 to 2000 MeV/nucleon.} 
\label{fig6}
\end{figure}

\begin{figure}[htp] 
\vspace{3.5cm}
\centering \epsfig{file=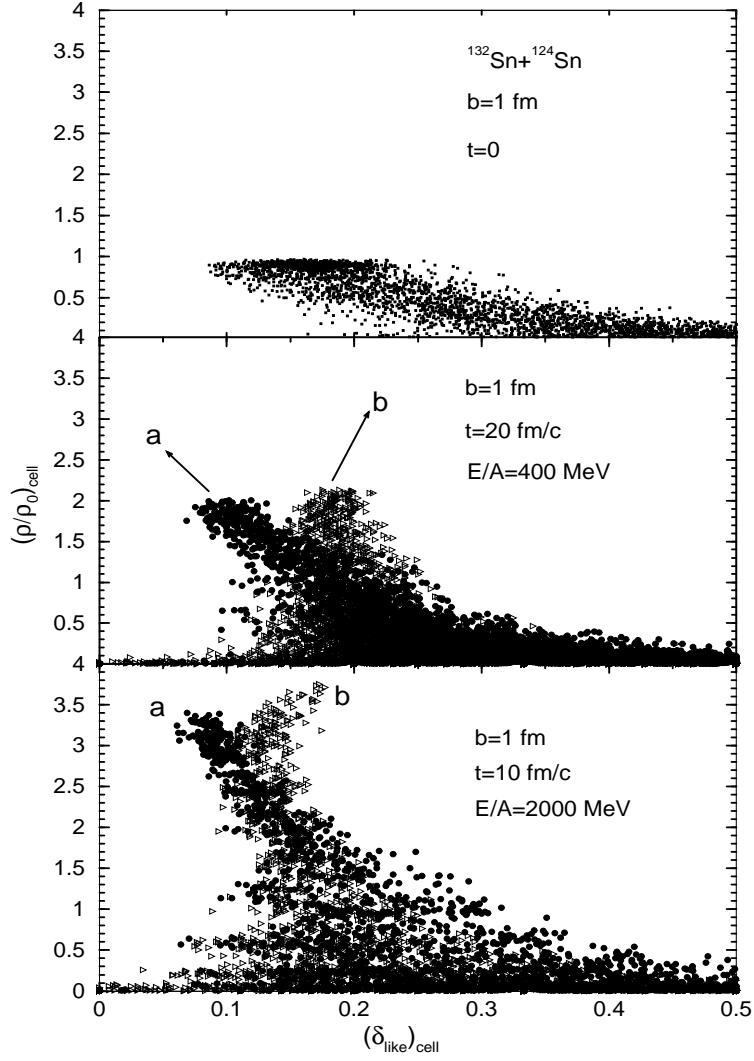,width=14cm,height=10cm,angle=-90} 
\vspace{1.cm}
\caption{Scatter plots of isospin asymmetry-density correlation 
in the initial state (upper window) and at the instants of maximum 
compressions in the reaction of $^{132}Sn+^{124}Sn$ with the
nuclear symmetry energy $E^a_{sym}$ (filled circles) and $E^b_{sym}$ 
(right triangles), respectively.}
\label{fig7}
\end{figure}

\begin{figure}[htp] 
\vspace{3.5cm}
\centering \epsfig{file=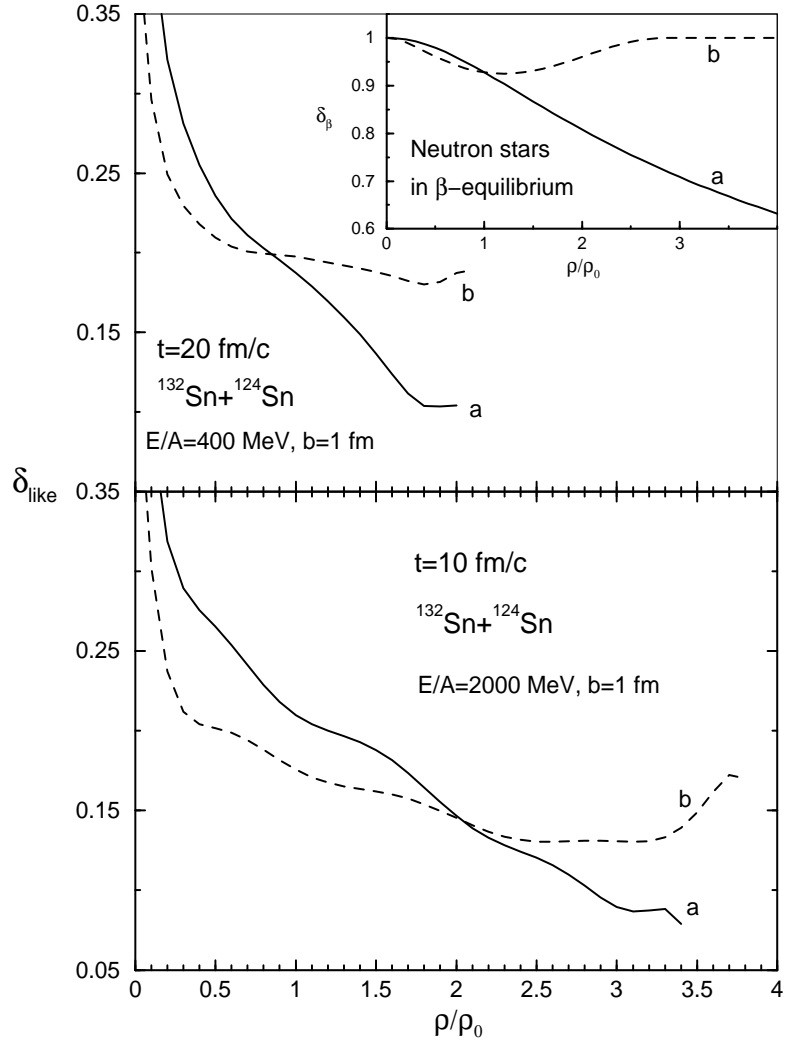,width=14cm,height=10cm,angle=-90} 
\vspace{1.cm}
\caption{Upper window: the isospin asymmetry-density correlations at t=20 fm/c
and $E_{beam}/A=400$ MeV in the central $^{132}Sn+^{124}Sn$ reaction with the
nuclear symmetry energy $E^a_{sym}$ and $E^b_{sym}$, respectively.
Lower window: the same correlation as in the upper window but 
at 10 fm/c and $E_{beam}/A=2$ GeV/nucleon. The corresponding correlation 
in neutron stars is shown in the insert.} 
\label{fig8}
\end{figure}

\begin{figure}[htp]
\vspace{3.5cm} 
\centering \epsfig{file=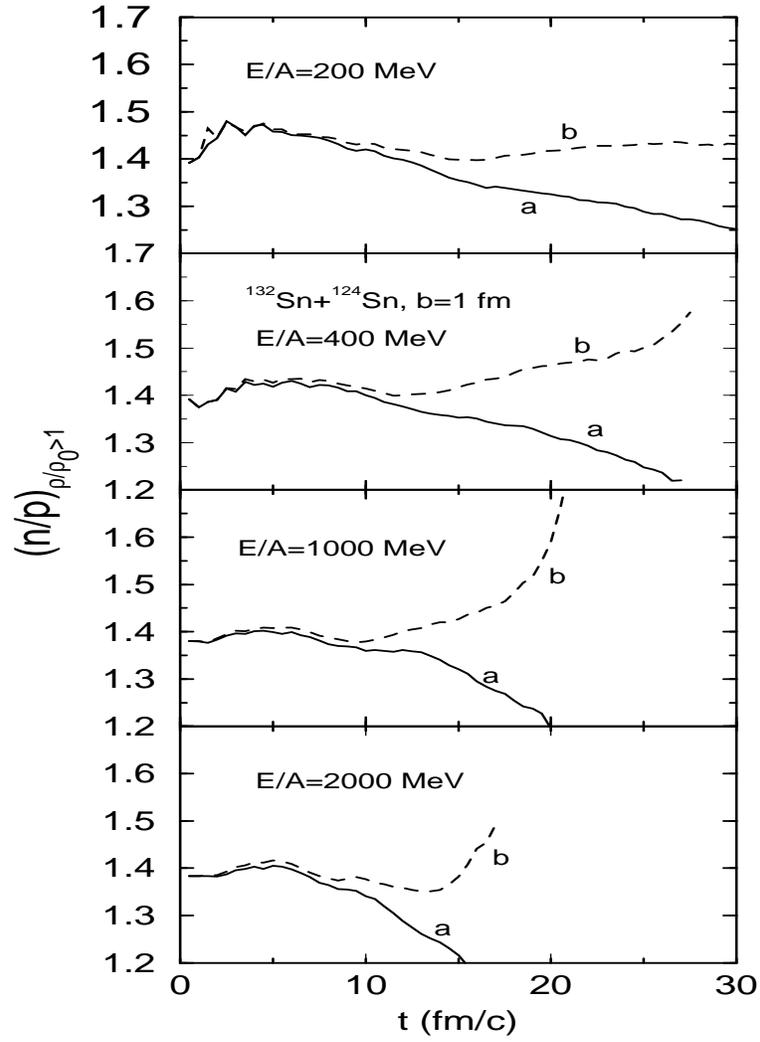,width=14cm,height=10cm,angle=-90} 
\vspace{1.cm}
\caption{The neutron/proton ratio of nuclear matter with density
higher than the normal nuclear matter density as a function of time
with the nuclear symmetry energy $E^a_{sym}$ and $E^b_{sym}$, respectively.} 
\label{fig9}
\end{figure}

\begin{figure}[htp]
\vspace{3.5cm} 
\centering \epsfig{file=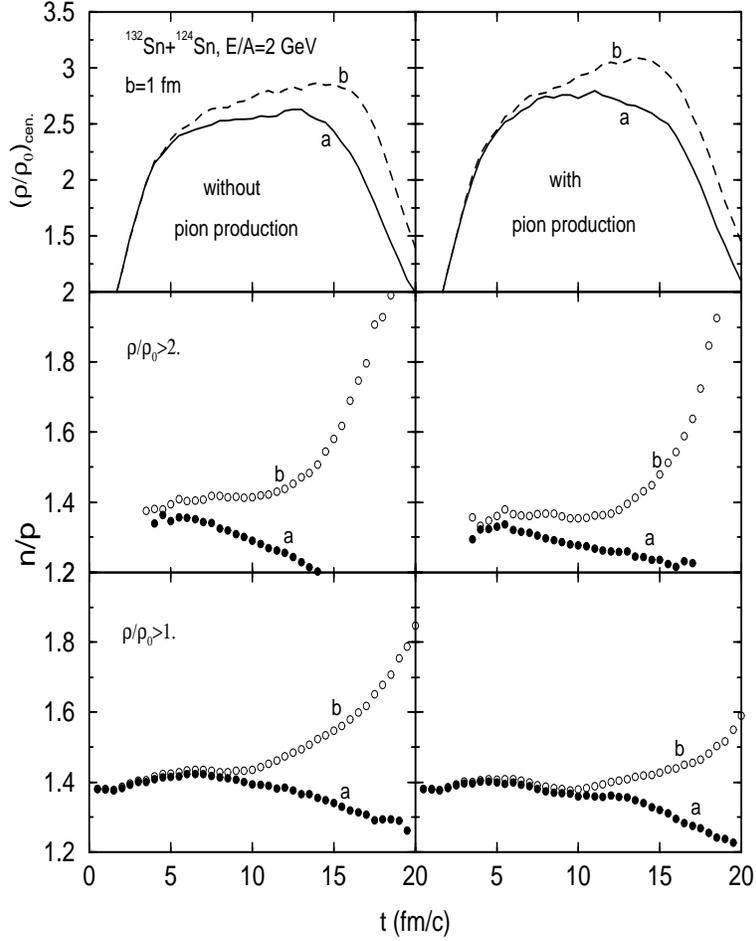,width=14cm,height=10cm,angle=-90} 
\vspace{1.cm}
\caption{Central baryon density (top row) and the neutron/proton ratio 
of nuclear matter with density $\rho\geq 2\rho_0$ (middle row) 
and $\rho\geq \rho_0$ (bottom row) with (right colomn) and without (left colomn)
the pion production channel in the central $^{132}Sn+^{124}Sn$ reaction at a beam energy of
2 GeV/nucleon.} 
\label{fig10}
\end{figure}

\begin{figure}[htp]
\vspace{3.5cm} 
\centering \epsfig{file=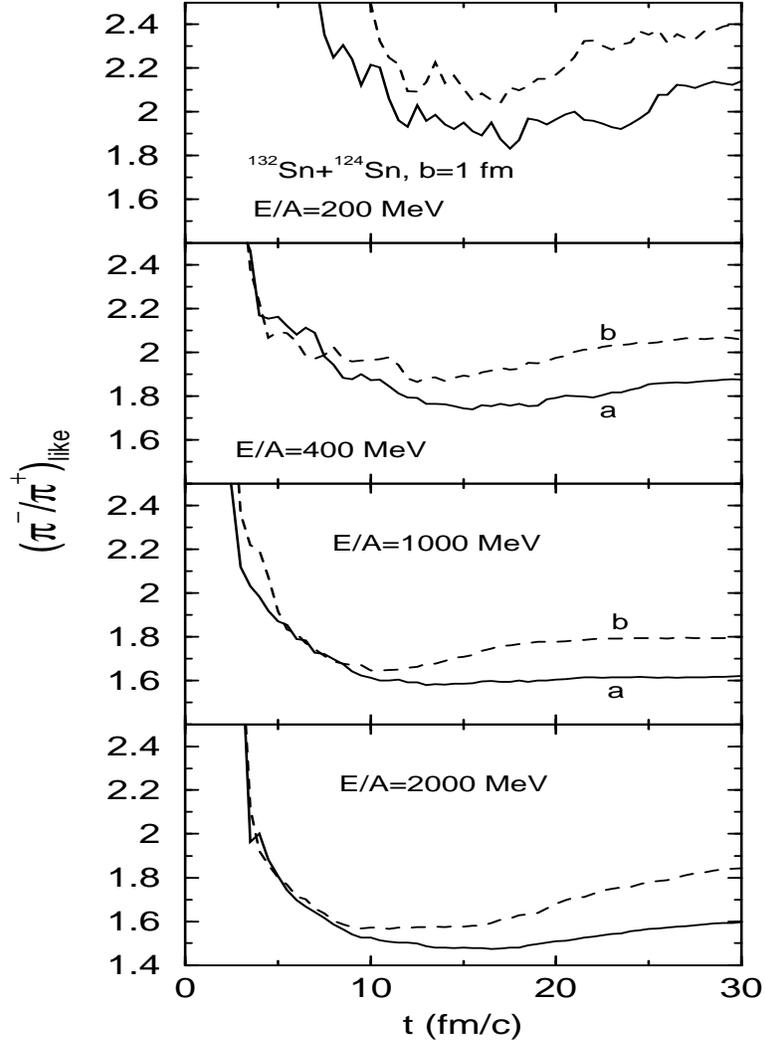,width=14cm,height=10cm,angle=-90} 
\vspace{1.cm}
\caption{ $\pi^-/\pi^+$ ratio as a function of time  
in the central $^{132}Sn+^{124}Sn$ reaction at a beam energy of
between 200 and 2000 MeV/nucleon.} 
\label{fig11}
\end{figure}

\begin{figure}[htp]
\vspace{3.5cm} 
\centering \epsfig{file=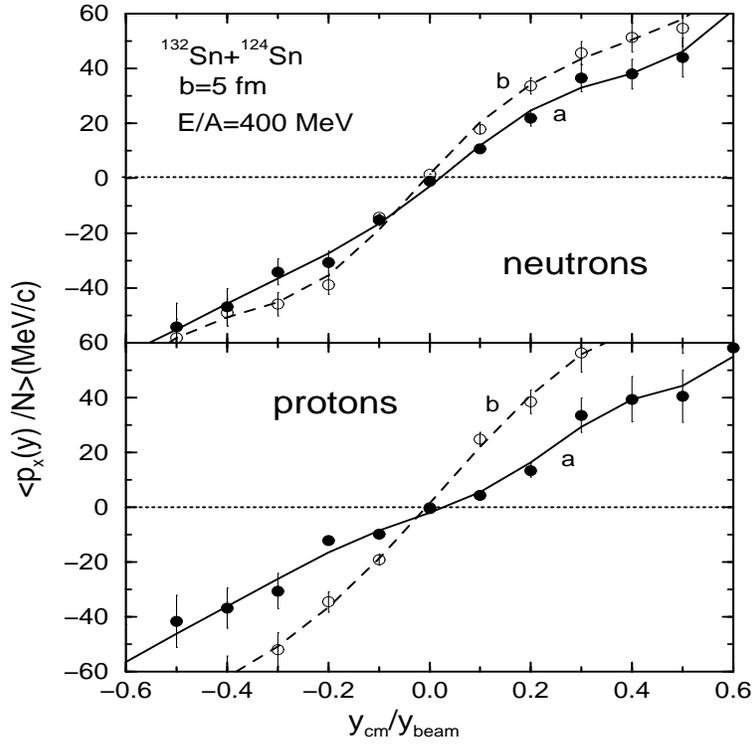,width=10cm,height=10cm,angle=-90} 
\vspace{1.cm}
\caption{ Transverse flow analysis for nuetrons (upper window) and
protons (lower window) in the mid-central $^{132}Sn+^{124}Sn$ 
reaction at a beam energy of 400 MeV/nucleon.} 
\label{fig12}
\end{figure}

\begin{figure}[htp]
\vspace{3.5cm} 
\centering \epsfig{file=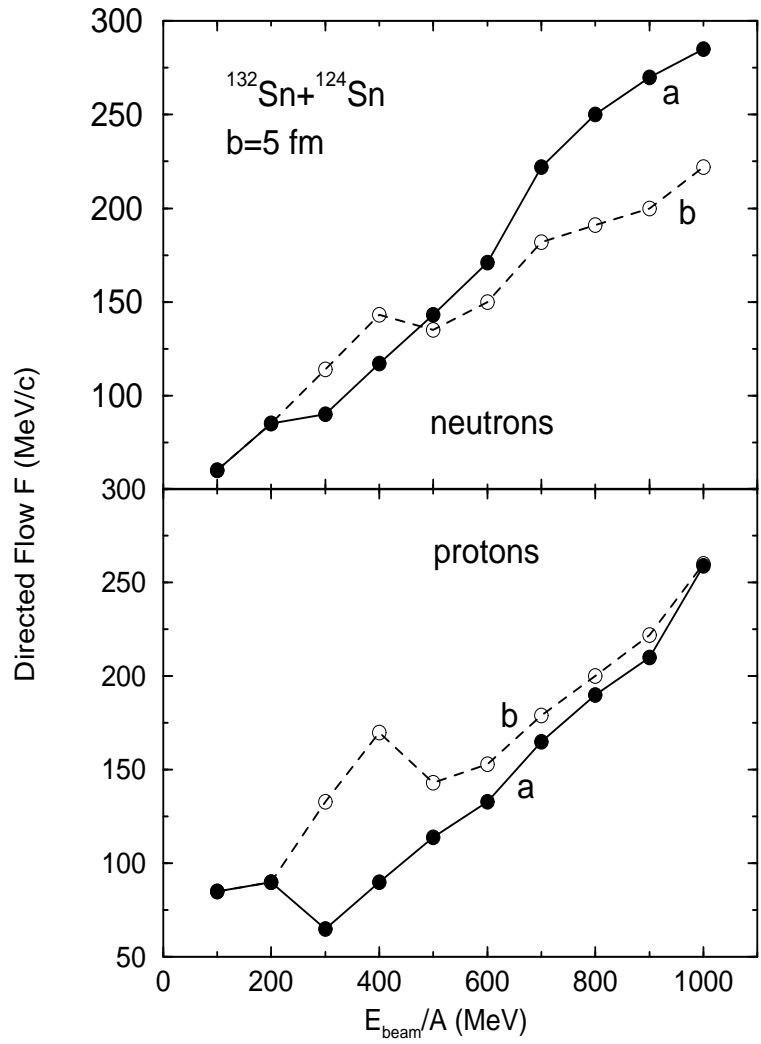,width=14cm,height=10cm,angle=-90} 
\vspace{1.cm}
\caption{The excitation functions of the transverse flow parameter for 
neutrons (upper) and protons (lower) in the mid-central $^{132}Sn+^{124}Sn$ 
reactions.} 
\label{fig13}
\end{figure}

\begin{figure}[htp] 
\vspace{3.5cm}
\centering \epsfig{file=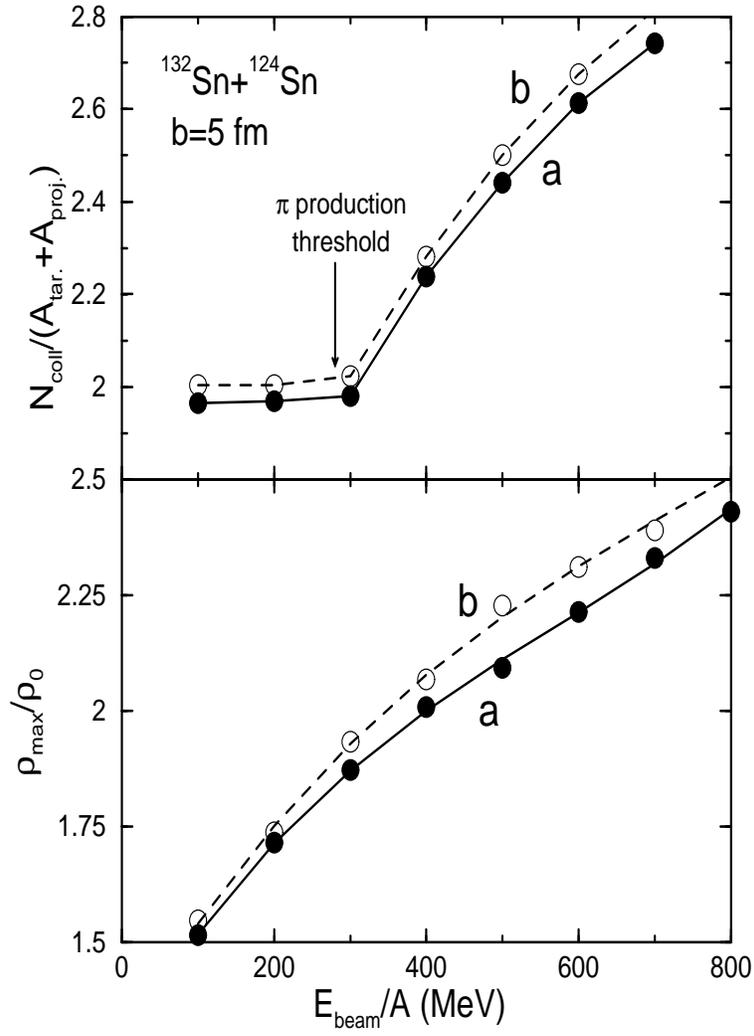,width=14cm,height=10cm,angle=-90} 
\vspace{1.cm}
\caption{The average number of successful hadronic scatterings 
per initial nucleon (upper) and the maximum baryon density reached in the 
mid-central $^{132}Sn+^{124}Sn$ reaction as a function of beam energy.}
\label{fig14}
\end{figure}

\begin{figure}[htp] 
\vspace{3.5cm}
\centering \epsfig{file=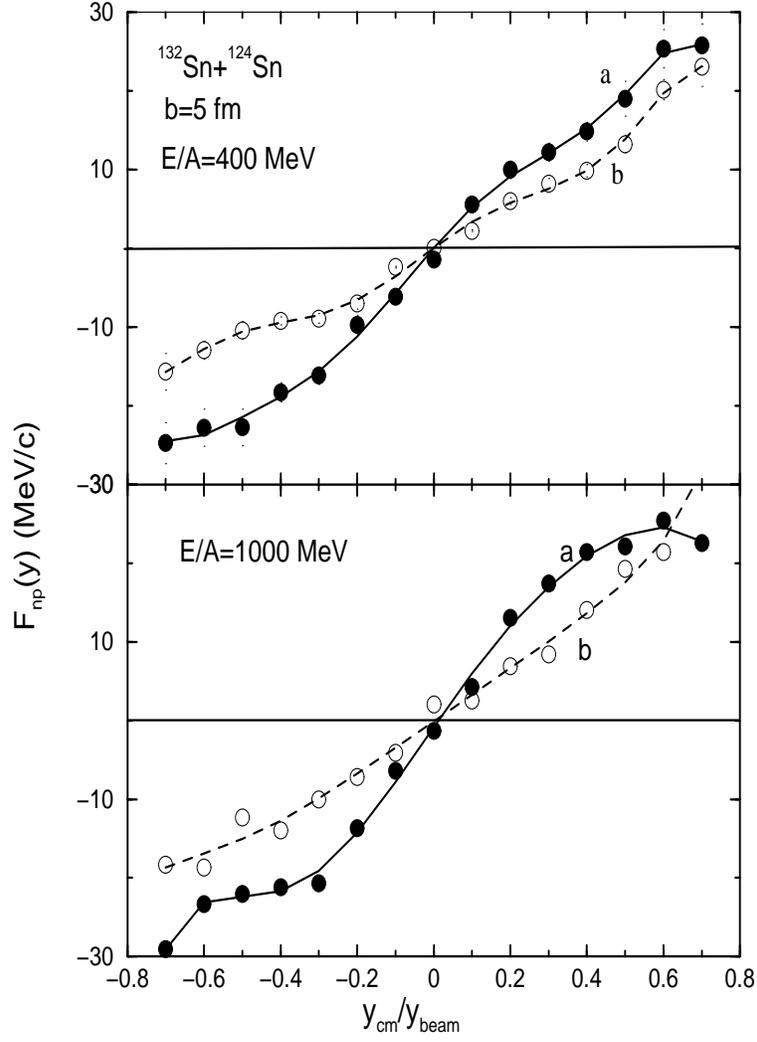,width=14cm,height=10cm,angle=-90} 
\vspace{1.cm}
\caption{The neutron-proton differential collective flow
in the mid-central $^{132}Sn+^{124}Sn$ reactions at $E_{beam}/A=400$ MeV (upper window) 
and 1000 MeV (lower window) with the nuclear symmetry energy 
$E^a_{sym}$ and $E^b_{sym}$, respectively.} 
\label{fig15}
\end{figure}

\begin{figure}[htp] 
\vspace{3.5cm}
\centering \epsfig{file=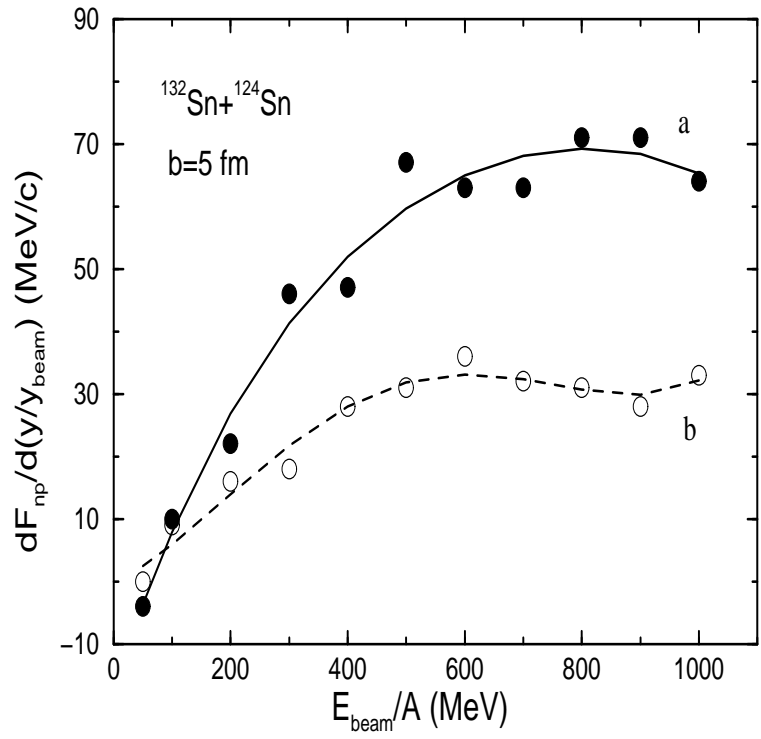,width=10cm,height=10cm,angle=-90} 
\vspace{1.cm}
\caption{Excitation function of the slope parameter of the neutron-proton 
differential flow for the mid-central $^{132}Sn+^{124}Sn$ reaction. The lines are drawn to 
guide the eye}
\label{fig16}
\end{figure}

\end{document}